\documentclass[12pt]{article}
\usepackage{amssymb}
\usepackage{graphics}
\usepackage{epsfig}
\usepackage{a4wide}

\begin{document}
\newcommand{\eewwz}{$e^+e^- \to W^+ W^-Z^0$ }
\newcommand{\eewwzg}{$e^+e^- \to W^+ W^-Z^0\gamma$ }

\title{ Full electroweak one-loop corrections to $W^+ W^-Z^0$ production at the ILC }
\author{ Sun Wei, Ma Wen-Gan, Zhang Ren-You, Guo Lei, and Song Mao \\
{\small Department of Modern Physics, University of Science and Technology }\\
{\small of China (USTC), Hefei, Anhui 230027, P.R.China}  }

\date{}
\maketitle \vskip 15mm
\begin{abstract}
The precise investigation of the $W^+ W^-Z^0$ production at the
$e^+e^-$ International Linear Collider(ILC) is of crucial
importance in probing the couplings between massive vector gauge
bosons and discovering the signature of new physics beyond the
standard model(SM). We study the full one-loop EW effects on the
observables, such as, the total cross section, the differential
cross section of the invariant mass of $W$-pair, the distribution
of the angle between $W$-pair, the production angle distributions
of $W$- and $Z^0$-boson, the distributions of the transverse
momenta of final $W$- and $Z^0$-boson, and the forward-backward
charge asymmetry of $W^-$-boson. Our numerical results show that
the EW relative correction to the total cross
sections($\delta_{ew}$) varies from $-17.6\%$ to $-5.3\%$ when
$m_H=120~GeV$ and $\sqrt{s}$ goes up from $300~GeV$ to $1~TeV$.
\end{abstract}

\vskip 2cm {\large{\bf Keywords}: $W^+ W^-Z^0$ production, W-boson
pair, electroweak radiative corrections } \\

{\large\bf PACS: 12.15.Lk, 12.38.Bx, 14.70.Hp, 14.70.Pw }

\vfill \eject

\baselineskip=0.32in

\renewcommand{\theequation}{\arabic{section}.\arabic{equation}}
\renewcommand{\thesection}{\Roman{section}.}
\newcommand{\nb}{\nonumber}

\newcommand{\Dir}{\kern -6.4pt\Big{/}}
\newcommand{\Dirin}{\kern -10.4pt\Big{/}\kern 4.4pt}
\newcommand{\DDir}{\kern -7.6pt\Big{/}}
\newcommand{\DGir}{\kern -6.0pt\Big{/}}

\makeatletter      
\@addtoreset{equation}{section}
\makeatother       

\section{Introduction}
\par
The Higgs mechanism plays an important role in the Standard Model
(SM) \cite{s1,s2}. It describes that the longitudinally polarized
components of the physical $Z^0$- and $W^{\pm}$-bosons eat the
hidden degrees of freedom of the Higgs field. The $SU(2)\times
U(1)$ gauge invariance provides stringent constraints on the
strengthes of triple and quartic gauge couplings. The accurate
measurements of these couplings could provide the information
about the electroweak (EW) symmetry breaking.

\par
The multiple gauge boson productions are suitable for probing the
self-coupling properties of the gauge bosons, and would give a
crucial test of the non-Abelian structure of the SM. If the
measured cross section is in agreement with the SM prediction, we
can put a severe constraint on new physics. On the contrary, if
there really exist gauge boson anomalous couplings, it would
generally lead to sizable effects on the EW observables.
Therefore, probing gauge couplings and searching for possible
anomalous contributions due to the effects of new physics is one
of the most important tasks of the present and future high energy
experiments.

\par
Among all the gauge boson self-couplings, the triple gauge
couplings(TGCs) of the neutral EW bosons $Z^0$, $\gamma$ and the
charged bosons $W^{\pm}$ have been well measured at the LEP2
\cite{measurement of charged TGC at LEP2}. The $e^+e^- \to W^+W^-$
process at the LEP2 was measured not only for determining the $W$
mass, but also for probing the charged TGCs\cite{hep-ex/0412015}.
To match the experimental accuracy, the one-loop level EW
corrections to $e^+e^- \to W^+W^-$ and $e^+e^- \to W^{+*}W^{-*}
\to 4f$ were calculated in Refs.\cite{part 1, part 2}. The
logarithmically enhanced two-loop electroweak radiative
corrections to the differential cross section for $W$-pair
production at the ILC up to the second power of the large
logarithm were also provided in Ref.\cite{2-loop corrections to
W-pair production}. The experiments at LEP2 demonstrated that the
SM expectations are in good agreement with the experimental data
within a few percent\cite{hep-ex/0412015}. If the colliding energy
is larger than the threshold of $Z^0$-boson pair production, the
$Z^0$-pair production process can be used to probe the neutral
TGCs. At the Fermilab Tevatron, the CDF and D0 collaborations
performed also some experiments about the diboson production in
$p\bar{p}$ collisions at $\sqrt{s}=1.96~TeV$, and presented the
limitations on anomalous TGCs in Ref.\cite{diboson production at
the Tevatron}.

\par
Triple massive gauge boson production processes, such as
$Z^0Z^0Z^0$ and $W^+W^-Z^0$ productions, will be investigated at
the Large Hadron Collider(LHC) and future International Linear
Collider(ILC). These processes can be used to probe the quartic
gauge couplings(QGCs). In Ref.\cite{VVV production at hadron
colliders}, the precise predictions for the $VVV$ productions at
hadron colliders were provided. It shows that the QCD corrections
increase the $Z^0Z^0Z^0$ and the $W^+W^-Z^0$ cross sections at the
LHC by about $50\%$ and $70\%$, respectively. Therefore, any
quantitative measurement of the concerned gauge couplings at
hadron colliders will have to take QCD corrections into account.

\par
Compared to hadron machine, $e^+e^-$ linear collider has the
advantage in performing experimental measurement with a
particularly clean environment. Actually, our present knowledge
about particle physics came from both types of colliders. For
example, the $Z^0$ and $W^\pm$ massive gauge bosons were firstly
discovered at a hadron collider, but their detailed properties and
roles in the SM theory were from the LEP experiments. Therefore,
lepton and hadron colliders are always complementary machines.

\par
The future ILC is an efficient machine for precise experiments
with $e^+e^-$ colliding energy range of $200~GeV<\sqrt{s}<500~GeV$
in the near future. It would be upgraded to $\sqrt{s}\sim
1~TeV$\cite{ILC}. This machine has sufficient energy to produce
multiple massive vector bosons, and would be ideally suited to
precision studies of the self-couplings of the vector gauge
bosons. For example, the reactions $e^+ e^- \to Z^0Z^0Z^0$ and
$e^+ e^- \to W^+W^-Z^0$ are very important processes at the ILC
for probing the quartic massive gauge couplings with high
precision. The process $e^+ e^- \to Z^0Z^0Z^0$ can be used to
provide some informations about the anomalous $Z^0Z^0Z^0Z^0$
coupling, and its one-loop EW corrections have been calculated in
Ref.\cite{sujijuan}. The phenomenology of the process $e^+ e^- \to
W^+W^-Z^0$ at the leading order (LO) was studied in
Ref.\cite{tree-level results for WWZ production}. In order to
match the experimental accuracy, it is necessary to take into
account the EW radiative corrections in the theoretical
predictions.

\par
In this work we calculate the complete one-loop EW corrections to
the process \eewwz in the SM. The paper is organized as follows:
In Section 2 we describe the calculations of the leading-order
(LO) cross section and the full ${\cal O}(\alpha_{ew})$ EW
radiative corrections to the \eewwz process. In Section 3 we
present some numerical results and discussion. Section 4
summarizes the conclusions.

\vskip 8mm
\section{Calculations}
\par
\begin{figure*}
\begin{center}
\includegraphics*[127pt,210pt][530pt,710pt]{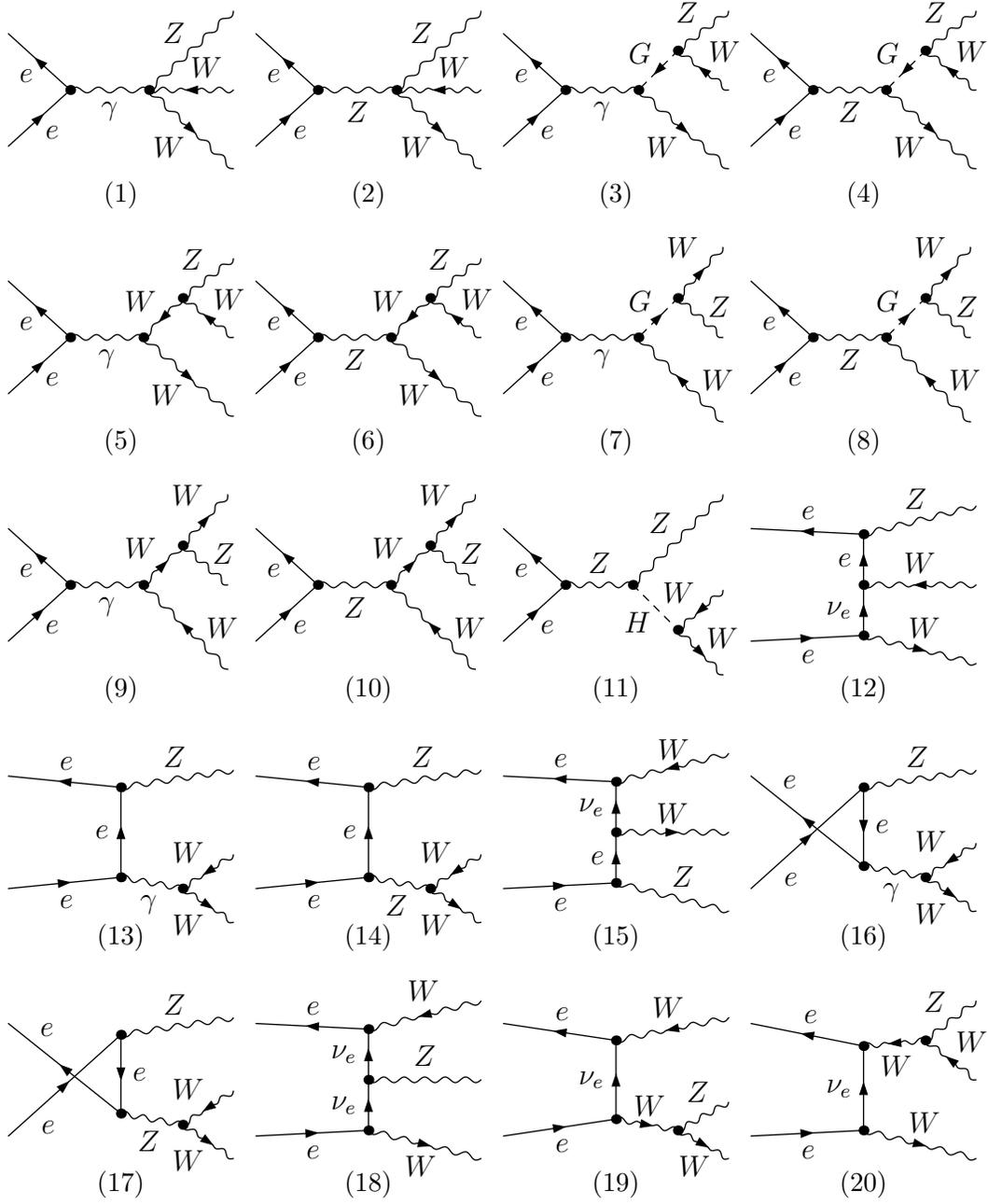}
\caption{\label{fig1} The tree-level Feynman diagrams for the \eewwz
process. }
\end{center}
\end{figure*}

We adopt the 't Hooft-Feynman gauge in the LO and next-to-leading
order(NLO) calculations, except when we verify the gauge
invariance at the LO. The FeynArts3.3 package\cite{fey} is
employed to generate the Feynman diagrams and their corresponding
amplitudes. The reductions of the amplitude are mainly implemented
by using FormCalc5.3 programs\cite{formloop}. Since the
contribution from the Feynman diagrams involving $H^0-e^+-e^-$ or
$G^0-e^+-e^-$ coupling is negligible due to the Yukawa coupling
strength being proportional to the related fermion mass, we do not
involve these graphs in our calculation. Then there are twenty
Feynman diagrams for the process \eewwz at the tree-level(shown in
Fig.\ref{fig1}). We denote the process \eewwz as
\begin{equation}
\label{process} e^+(p_1)+e^-(p_2) \to W^+(p_3)+W^-(p_4)+Z^0(p_5),
\end{equation}
The differential cross section for the process \eewwz at the LO is
then obtained as
\begin{eqnarray} \label{cross}
d\sigma_{tree} =\frac{1}{4} \sum_{spin}|{\cal M}_{tree}|^2 d\Phi_3 ,
\end{eqnarray}
where ${\cal M}_{tree}$ is the amplitude of all the tree-level
diagrams, and the factor $\frac{1}{4}$ is from taking average over
the spins of the initial particles. The three-particle phase space
element $d\Phi_3$ is defined as
\begin{eqnarray}
d\Phi_3=\delta^{(4)} \left( p_1+p_2-\sum_{i=3}^5 p_i \right)
\prod_{j=3}^5 \frac{d^3 \textbf{\textsl{p}}_j}{(2 \pi)^3 2 E_j}.
\end{eqnarray}

\par
\begin{figure}[htbp]
\vspace*{-0.3cm} \centering
\includegraphics*[130pt,410pt][530pt,710pt]{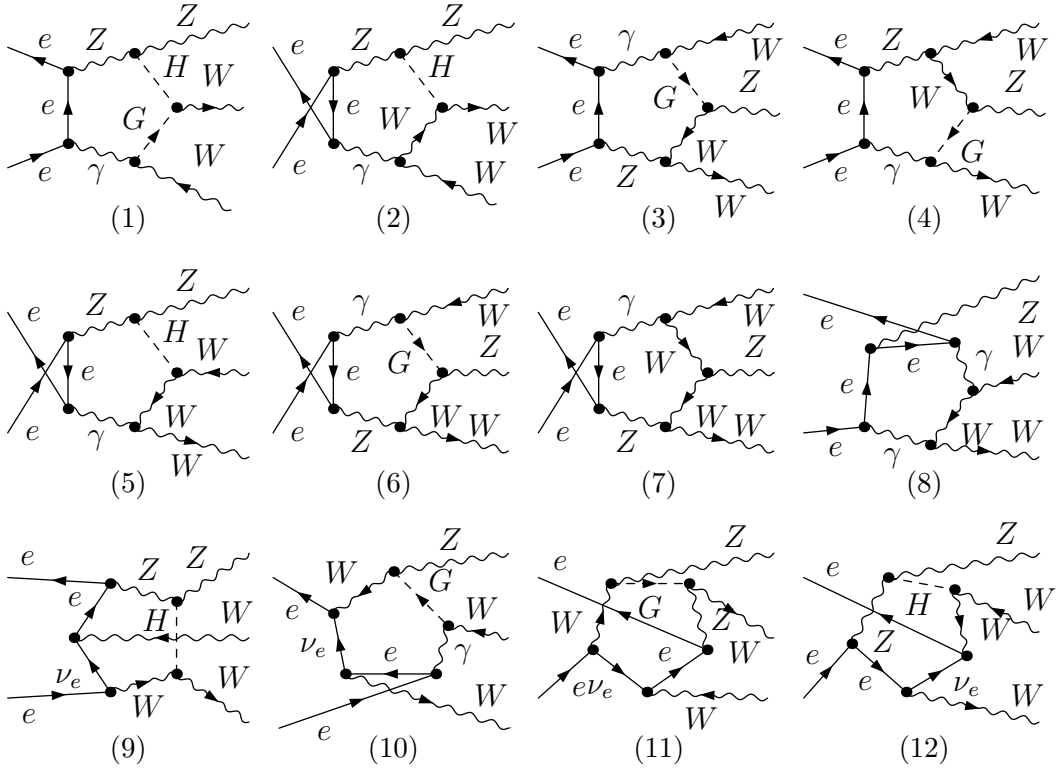}
\vspace*{-0.3cm} \centering \caption{\label{fig2} Some
representative pentagon Feynman diagrams for the process \eewwz. }
\end{figure}

In the EW NLO calculation we take the definitions of one-loop
integral functions as presented in Ref.\cite{Denner}. The complete
EW one-loop Feynman diagrams include 3510 graphs, and we organize
them into self-energy(1280), triangle(1357), box(605), pentagon(140)
and counterterm(128) diagram groups. Some of the pentagon graphs are
depicted in Fig.\ref{fig2} as a representative selection. We adopt
the dimensional regularization(DR) scheme\cite{DR} to regularize all
the soft IR and UV divergencies, where the dimensions of spinor and
space-time manifolds are extended to $D = 4 - 2 \epsilon$, to
isolate the UV and IR divergences. The collinear IR singularities
are regularized by keeping finite electron/positron mass. The
Cabibbo-Kobayashi-Maskawa(CKM) matrix is assumed to be identity
matrix in our calculation. We adopt the definitions for the relevant
renormalization constants as presented in Ref.\cite{Denner}. Using
the on-mass-shell conditions\cite{COMS scheme}, the relevant
renormalized constants can be expressed as\cite{Denner}
\begin{eqnarray}
\delta m_e=\frac{m_e}{2}\tilde{Re}\left[ \Sigma_e^L(m_e^2)+
\Sigma_e^R(m_e^2)+2\Sigma_e^S(m_e^2) \right],~~~
\delta m_{Z}^2=\tilde{Re}\Sigma_T^{ZZ}(m_{Z}^2),~~~\nb \\
\delta Z_e^{L(R)}=-\tilde{Re} \Sigma_e^{L(R)}(m_e^2)-m_e^2
\frac{\partial}{\partial p^2}\tilde{Re}\left[
\Sigma_e^L(p^2)+\Sigma_e^R(p^2)+2\Sigma_e^S(p^2) \right]|_{p^2=m_e^2},~~~\nb \\
\delta m_{W}^2=\tilde{Re}\Sigma_T^{W}(m_{W}^2),~~~\delta
Z_{W^{\pm}}=-\tilde{Re}\frac{\partial \Sigma^W(p^2)}{\partial
p^2}|_{p^2=m_W^2},~~~\delta Z_{AA}=- \frac{\partial
\Sigma_T^{AA}(p^2)}{\partial
p^2}|_{p^2=0},~~~\nb \\
\delta Z_{ZZ}= - \tilde{Re} \frac{\partial
\Sigma_T^{ZZ}(p^2)}{\partial p^2}|_{p^2=m_Z^2}, ~~~\delta
Z_{ZA}=2\frac{\Sigma_T^{AZ}(0)}{m_Z^2}, ~~~ \delta Z_{AZ}= - 2 Re
\frac{\Sigma_T^{AZ}(m_Z^2)}{m_Z^2}.
\end{eqnarray}
For the derived charge renormalization constant and the
counterterm of the parameter $s_W$, we have\cite{Denner}
\begin{eqnarray}
\delta Z_e=-\frac{1}{2}\delta
Z_{AA}-\frac{s_W}{c_W}\frac{1}{2}\delta Z_{ZA},~~~\frac{\delta
s_W}{s_W}=-\frac{1}{2}\frac{c_W^2}{s_W^2}\tilde{Re}\left(
\frac{\Sigma_T^W(m_W^2)}{m_W^2}-\frac{\Sigma_T^{ZZ}(m_Z^2)}{m_Z^2}\right).
\end{eqnarray}

\par
The reductions of the vector and tensor integrals are done exactly
by using the approach presented in Refs.\cite{Passarino,Five}. The
numerical calculations of the scalar one-, two-, three-, four- and
five-point integral functions are processed according to the
expressions presented in Refs.\cite{Five,OneTwoThree,Four}. The
calculations are carried out by using LoopTools-2.4
package\cite{formloop}\cite{van} and our independently developed
programs for the calculations of scalar, vector and tensor
five-point integrals with the approach presented in Ref.\cite{Five}
separately, in order to cross check for possible numerical
instabilities. The virtual contribution of ${\cal O}(\alpha_{ew}^4)$
to \eewwz process can be expressed as\cite{hepdata},
\begin{eqnarray}
\Delta\sigma_{{\rm vir}} = \sigma_{{\rm tree}} \delta_{{\rm vir}} =
\frac{(2 \pi)^4}{2 |\vec{p}_1| \sqrt{s}} \int \frac{1}{4}{\rm d}
\Phi_3 \sum_{{\rm spin}} {\rm Re} \left( {\cal M}_{{\rm tree}} {\cal
M}_{{\rm vir}}^{\dag} \right),
\end{eqnarray}
where $\vec{p}_1$ is the c.m.s. spatial momentum of the incoming
positron. ${\cal M}_{{\rm vir}}$ represents the renormalized
amplitude of one-loop Feynman diagrams.

\par
According to the Kinoshita-Lee-Nauenberg (KLN) theorem\cite{KLN},
we should consider the contribution of the real photon emission
process \eewwzg in order to get the IR safe observables for the
process \eewwz at the NLO. There includes 148 tree-level Fynman
diagrams for the photon emission process \eewwzg. In the
calculation of this process, we adopt the phase-space-slicing
(PSS) method \cite{PSS} to isolate the soft photon emission
singularity. We divide the photon phase space into two parts: If
$E_{\gamma} \leq \delta_s E_b$, it's called soft photon region. If
$E_{\gamma} > \delta_s E_b$, it's in hard photon region. Then the
cross section of the \eewwzg process can be expressed as
\begin{equation}
\Delta\sigma_{{real}}=\Delta\sigma_{{soft}}+\Delta\sigma_{{hard}}=
\sigma_{tree}(\delta_{{soft}}+\delta_{{hard}}),
\end{equation}
where only the term $\Delta\sigma_{{soft}}$ includes soft IR
singularity. Theoretically, both $\Delta\sigma_{{soft}}$ and
$\Delta\sigma_{{hard}}$ should depend on the arbitrary soft cutoff
$\delta_s$, but the total EW one-loop
correction($\Delta\sigma_{tot}$) and $\Delta\sigma_{{real}}$
should be cutoff $\delta_s$ independent.

\par
In dealing with the soft IR divergencies, we introduce a fictitious
small photon mass($m_{\gamma}$) for the internal photon lines of
loop diagrams, and reproduce the soft IR divergent integrals upon
the replacements of
\begin{eqnarray}\label{IR}
\pi^{-\epsilon_{IR}}\Gamma(\epsilon_{IR}) \to \ln
(m_{\gamma}^2),~~~~~
\pi^{-\epsilon_{IR}}\frac{\Gamma(\epsilon_{IR})}{\epsilon_{IR}}
\to \frac{1}{2} \ln^2 (m_{\gamma}^2),
\end{eqnarray}
where $m_{\gamma}$ is chosen with a sufficiently small value, but
not too small to induce numerical instabilities. After doing the
replacements of (\ref{IR}) for the IR divergent integrals, we give
up the use of DR scheme for the case of IR divergences and adopt the
massive photon scheme with a fictitious photon mass as regulator.
That replacements are also done in treating with the soft IR
singularity for the process \eewwzg before integrating over the
phase space for the emitted photon. Generally we take a small value
for $\delta_s$ in the calculation for the process \eewwzg. The terms
of order $\delta_s$ in $\Delta\sigma_{{soft}}$ can be neglected and
the $\Delta\sigma_{{soft}}$ can be evaluated analytically by fixing
a small photon mass value. In the hard photon phase space region,
$\Delta\sigma_{{hard}}$ is calculated with photon mass being set to
zero. After regularizing the soft IR divergencies with massive
photon scheme, the UV finiteness of the whole contributions from the
virtual one-loop diagrams and counterterms has been checked both
analytically and numerically in DR scheme.

\par
If the IR singularity in the soft photon emission process is
really cancelled by the virtual photonic corrections, the
independence of $\Delta\sigma_{tot}(\equiv \Delta\sigma_{virtual}
+\Delta\sigma_{real})$ on the cutoff $\delta_s$ and fictitious
photon mass $m_{\gamma}$, should be demonstrated in numerical
calculation. The phase space integration for hard photon emission
process \eewwzg can be computed directly by using the Monte Carlo
method, because it is UV and IR finite. In practice we perform
this integration in hard photonic region by using our in-house $2
\to 4$ integration program based on Monte Carlo integrator Vegas.
Finally, the EW NLO corrected total cross section($\sigma_{tot}$)
up to the order of ${\cal O}(\alpha^4_{ew})$ for the \eewwz
process is obtained by summing the ${\cal O}(\alpha^3_{ew})$ Born
cross section($\sigma_{tree}$), the ${\cal O}(\alpha^4_{ew})$
virtual cross section($\Delta\sigma_{vir}$), and the ${\cal
O}(\alpha^4_{ew})$ cross section of the real photon emission
process \eewwzg($\Delta\sigma_{real}$).
\begin{eqnarray}
\sigma_{{tot}}=\sigma_{{tree}} + \Delta\sigma_{{tot}}=
\sigma_{{tree}} + \Delta\sigma_{{vir}} + \Delta\sigma_{{real}} =
\sigma_{{tree}} \left( 1 + \delta_{ew} \right),
\end{eqnarray}
where $\delta_{ew}$ is the full ${\cal O}(\alpha_{ew})$ EW
relative correction.

\vskip 8mm
\section{Numerical results and discussion}
\par
In the following we perform the numerical evaluations at the LO
and EW NLO in the $\alpha_{ew}$-scheme and the relevant input
parameters are taken as\cite{hepdata}:
\begin{equation}
\begin{array}{lll}\label{input parameter}
m_Z=91.1876~GeV, & m_W=80.398~GeV, & \sin^2 \theta_W=1-\frac{m_W^2}{m_Z^2}=0.222646, \\
m_u=m_d=66~MeV,  & m_s=104~MeV, & m_c=1.27~GeV, \\
m_b=4.2~GeV, & m_t=171.2~GeV,  & m_e=0.510998910~keV,   \\
m_{\mu}=105.658389~MeV, & m_{\tau} = 1776.84~MeV, &
\end{array}
\end{equation}
where we use the effective values of the light quark masses ($m_u$
and $m_d$) which can reproduce the hadron contribution to the
shift in the fine structure constant
$\alpha_{ew}(m_Z^2)$\cite{leger}. In the LO and NLO calculations
we take the fine structure constant $\alpha_{ew}(0)=1/137.036$ as
input parameter.

\par
For the numerical verification for the correctness of our LO
calculation, we use both CompHEP-4.4p3 and FeynArts3.3/FormCalc5.3
packages to calculate the LO cross section of process \eewwz by
adopting 't Hooft-Feynman gauge and unitary gauge separately. The
numerical results are listed in Table \ref{tab1}. It shows they
are in good agreement.
\begin{table}
\begin{center}
\begin{tabular}{|c|c|c|c|}
\hline   $\sigma_{tree}(fb)$(FeynArts) &
$\sigma_{tree}(fb)$(FeynArts) & $\sigma_{tree}(fb)$ (CompHEP)
& $\sigma_{tree}(fb)$ (CompHEP)  \\
\hline  Feynman gauge & Unitary gauge & Feynman gauge & Unitary gauge  \\
\hline  35.810(4)  & 35.810(4) & 35.80(2)  & 35.80(2)  \\
\hline
\end{tabular}
\end{center}
\begin{center}
\begin{minipage}{15cm}
\caption{\label{tab1} The comparison of the numerical results of the
LO cross sections $\sigma_{tree}$ for \eewwz process with
$\sqrt{s}=500~GeV$ in Feynman gauge and unitary gauge, by taking the
related input parameters as in Eqs.(\ref{input parameter}) and using
FeynArts3.3/FormCalc5.3 and CompHEP-4.4p3 packages separately. }
\end{minipage}
\end{center}
\end{table}
\par
\begin{figure}
\includegraphics[scale=0.75]{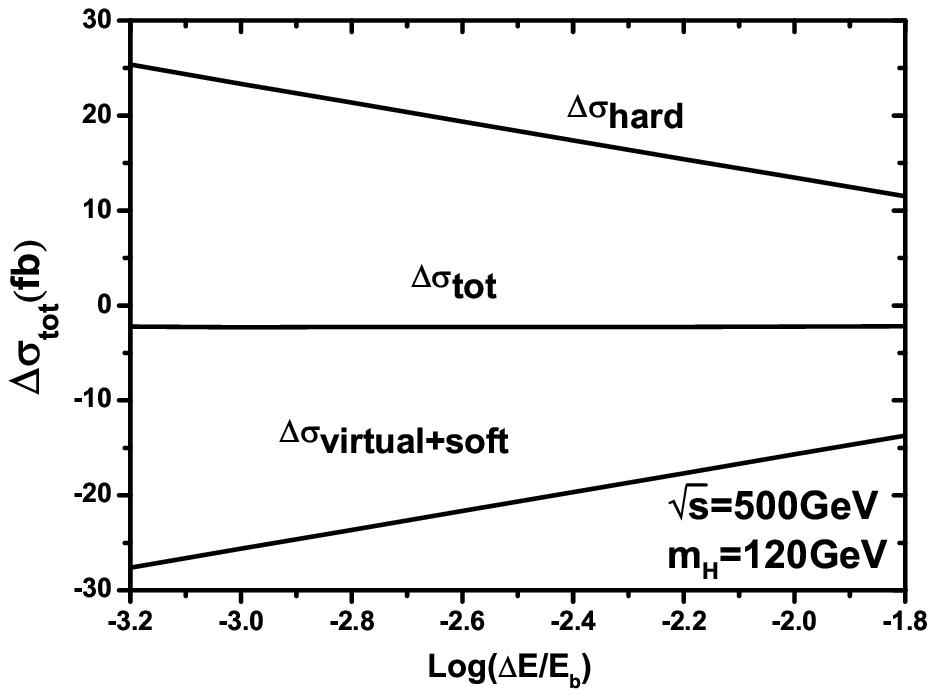}
\includegraphics[scale=0.75]{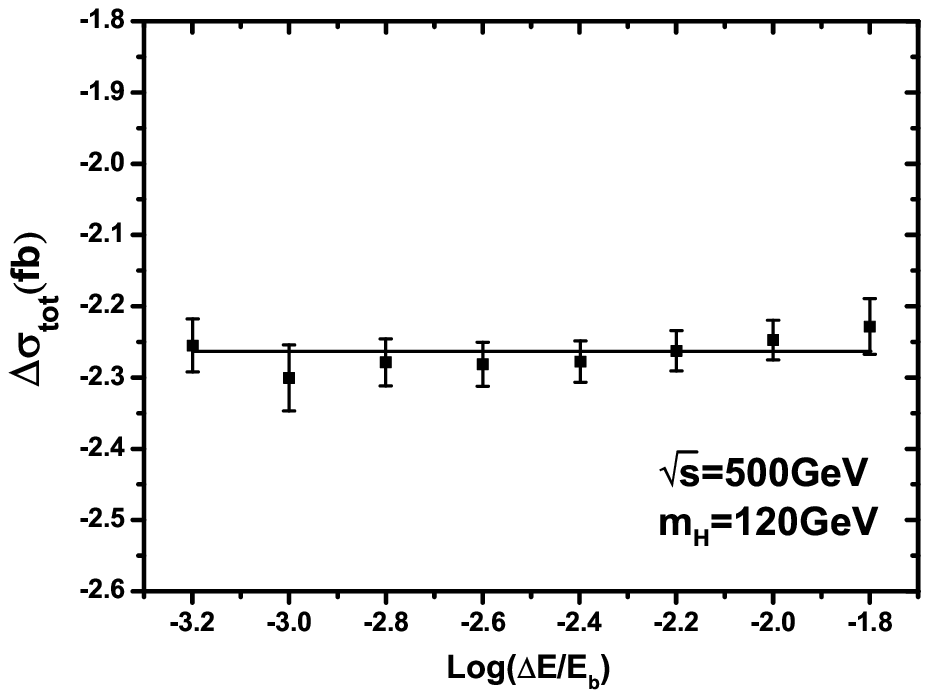}
\caption{\label{fig3} (a) The ${\cal O}(\alpha_{{\rm ew}}^4)$
contribution parts of cross section for \eewwz process as the
functions of the soft cutoff $\delta_s$ in conditions of
$m_{\gamma}=10^{-5}~GeV$, $m_H=120~GeV$ and $\sqrt{s}=500~GeV$.
(b) The amplified curve for $\Delta\sigma_{tot}$ of Fig.3(a)
versus $\delta_s$ including calculation errors.}
\end{figure}

As we mentioned in above section if our NLO calculation is correct
and the IR divergency is really cancelled, the total cross section
should be independent of $m_{\gamma}$ and $\delta_s$. In fact, our
calculation shows when the fictitious photon mass $m_{\gamma}$
varies from $10^{-15}~GeV$ to $10^{-1}~GeV$ in conditions of
$\delta_s=10^{-3}$, $m_H=120~GeV$ and $\sqrt{s}=500~GeV$, the
numerical results for the cross section correction
$\Delta\sigma_{tot}=\Delta\sigma_{real}+ \Delta\sigma_{vir}$, are in
mutual agreement up to ten effective digits. The independence of the
total EW NLO contribution to \eewwz process on soft cutoff
$\delta_s$ is demonstrated in Figs.\ref{fig3}(a,b), where we take
$m_{\gamma}=10^{-5}~GeV$, $m_H=120~GeV$ and $\sqrt{s}=500~GeV$. The
amplified curve for $\Delta\sigma_{tot}$ in Fig.\ref{fig3}(a) is
depicted in Fig.\ref{fig3}(b) including calculation errors.
Figs.\ref{fig3}(a,b) show that although both $\Delta\sigma_{{\rm
vir}}+\Delta\sigma_{{\rm soft}}$ and $\Delta\sigma_{{\rm hard}}$ are
strongly related to soft cutoff $\delta_s$, the total EW NLO
contribution $\Delta\sigma_{tot}=\Delta\sigma_{{\rm
vir}}+\Delta\sigma_{{\rm real}}$ is independent of the cutoff
$\delta_s$ within the range of calculation errors as expected. In
further calculations, we fix $m_{\gamma}=10^{-5}~GeV$ and
$\delta_s=10^{-3}$.

\par
\begin{figure}
\centering
\includegraphics[scale=0.75]{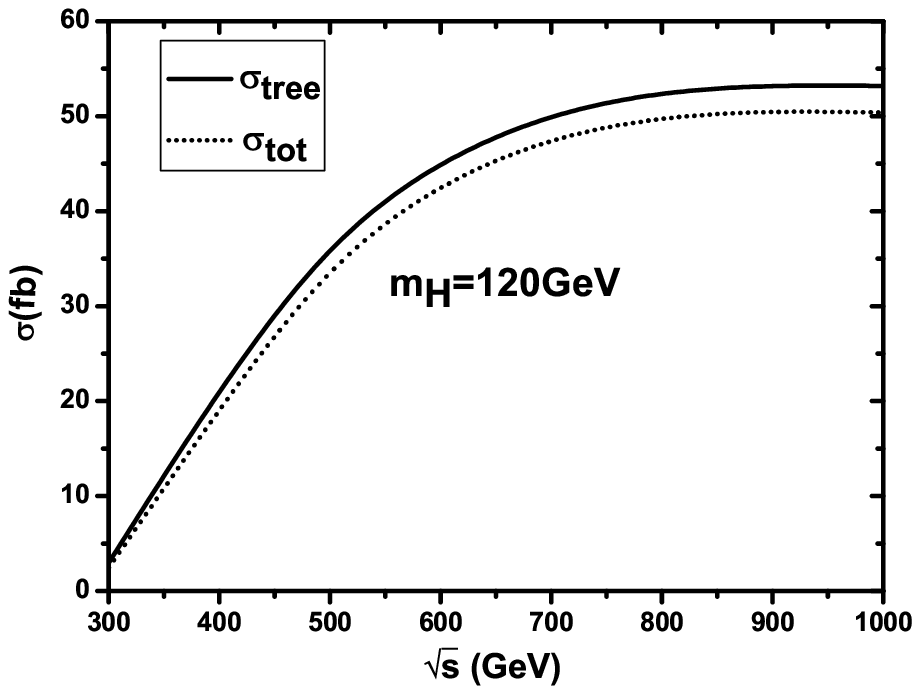}
\includegraphics[scale=0.75]{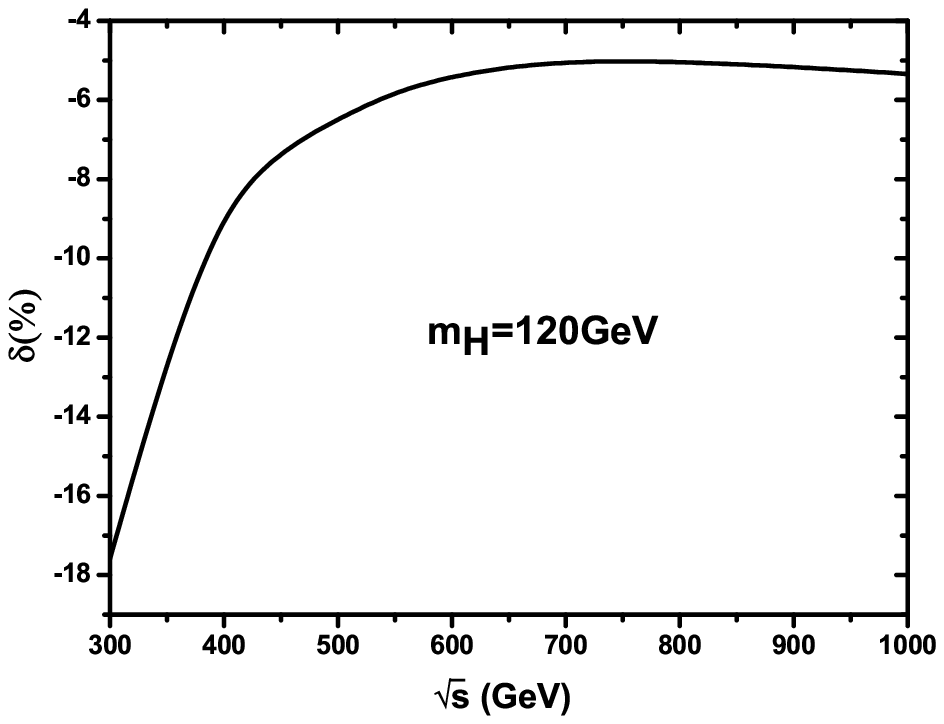}
\caption{\label{fig4} (a) The LO and EW NLO corrected cross
sections for the process \eewwz as the functions of colliding
energy $\sqrt{s}$ with $m_H=120~GeV$. (b) The corresponding EW
relative radiative corrections versus $\sqrt{s}$.}
\end{figure}

In Fig.\ref{fig4}(a) we depict the curves for the LO and EW NLO
corrected cross sections as the functions of colliding energy
$\sqrt{s}$ with $m_H=120~GeV$. Fig.\ref{fig4}(b) shows the
corresponding relative corrections($\delta_{ew}\equiv \frac{
\Delta\sigma_{tot}}{\sigma_{tree}}$) for the data drawn in
Fig.\ref{fig4}(a). We find from Figs.\ref{fig4}(a,b) that the LO
and EW NLO corrected cross sections are sensitive to the colliding
energy $\sqrt{s}$ in the range of $\sqrt{s}< 800~GeV$, and the LO
cross sections are suppressed by the EW NLO corrections in the
whole $\sqrt{s}$ range plotted in Fig.\ref{fig4}(a).
Fig.\ref{fig4}(b) shows that the absolute relative correction can
be very large in the vicinity where $\sqrt{s}$ approaches to the
threshold of $W^+W^-Z^0$ production. That effect comes from the
Coulomb singularity in the Feynman graphs involving the
instantaneous virtual photon exchange in loop which has a small
spatial momentum. To show the numerical results more exactly, we
list some representative numerical results of the LO, EW NLO
corrected cross sections($\sigma_{tree}$, $\sigma_{tot}$), the EW
NLO correction to the cross section
($\Delta\sigma_{tot}\equiv\sigma_{tot}-\sigma_{tree}$) and the EW
relative correction($\delta_{ew}\equiv
\Delta\sigma_{tot}/\sigma_{tree}$) in Table \ref{tab2}. There we
take $m_H=120~GeV$, $150~GeV$ and $\sqrt{s}=300~GeV$, $500~GeV$,
$800~GeV$, $1000~GeV$, separately. The results in Table \ref{tab2}
show that both the LO and NLO corrected cross sections for the
\eewwz process are insensitive to Higgs boson mass. From
Fig.\ref{fig4}(b) and Table \ref{tab2} we can see when
$m_H=120~GeV$ and $\sqrt{s}$ goes up from $300~GeV$ to $1~TeV$,
the EW relative radiative correction $\delta_{ew}$ varies from
$-17.6\%$ to $-5.3\%$.

\begin{table}
\begin{center}
\begin{tabular}{|c|c|c|c|c|c|}
\hline $\sqrt{s}$ (GeV) & $m_H(GeV)$& $\sigma_{tree}(fb)$ &
$\sigma_{tot}(fb)$ & $\Delta\sigma_{tot}(fb)$ & $\delta_{ew}($\%$)$  \\
\hline  300  & 120 & 2.9457(2)  & 2.427(2)  & -0.519(2)  & -17.62(7) \\
\hline  300  & 150 & 3.1605(2)  & 2.633(2)  & -0.527(2)  & -16.67(6) \\
\hline  500  & 120 & 35.810(4)  & 33.51(5)  & -2.30(5)   & -6.4(1)   \\
\hline  500  & 150 & 36.035(4)  & 33.85(5)  & -2.19(5)   & -6.1(1)   \\
\hline  800  & 120 &  52.34(1)  & 49.70(6)  & -2.64(5)   & -5.0(1)   \\
\hline  800  & 150 &  52.46(1)  & 50.10(6)  & -2.36(6)   & -4.5(1)   \\
\hline 1000  & 120 &  53.21(1)  & 50.37(7)  & -2.84(7)   & -5.3(1)   \\
\hline 1000  & 150 &  53.28(1)  & 50.70(7)  & -2.58(7)   & -4.8(1)   \\
\hline
\end{tabular}
\end{center}
\begin{center}
\begin{minipage}{15cm}
\caption{\label{tab2} The numerical results of the LO, EW NLO
corrected cross sections, the EW NLO correction to the cross
section ($\Delta\sigma_{tot}$) and the EW relative correction
($\delta_{ew}$) for the process \eewwz, in conditions of
$m_H=120~GeV$, $150~GeV$ and $\sqrt{s}=300~GeV$, $500~GeV$,
$800~GeV$, $1000~GeV$, separately. }\end{minipage}
\end{center}
\end{table}

\par
\begin{figure}
\centering
\includegraphics[scale=0.75]{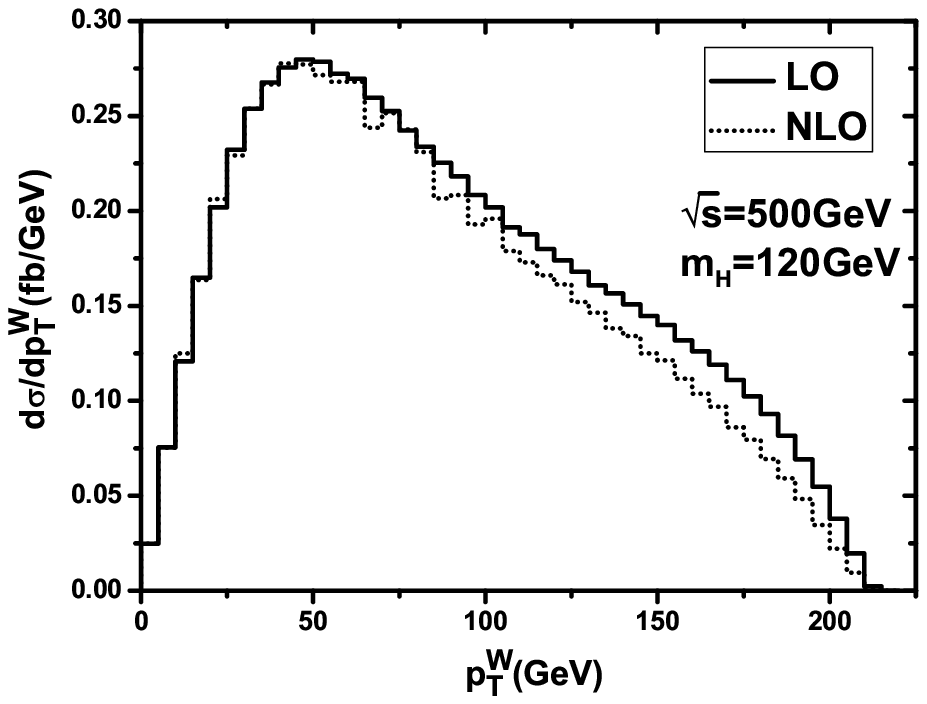}
\includegraphics[scale=0.75]{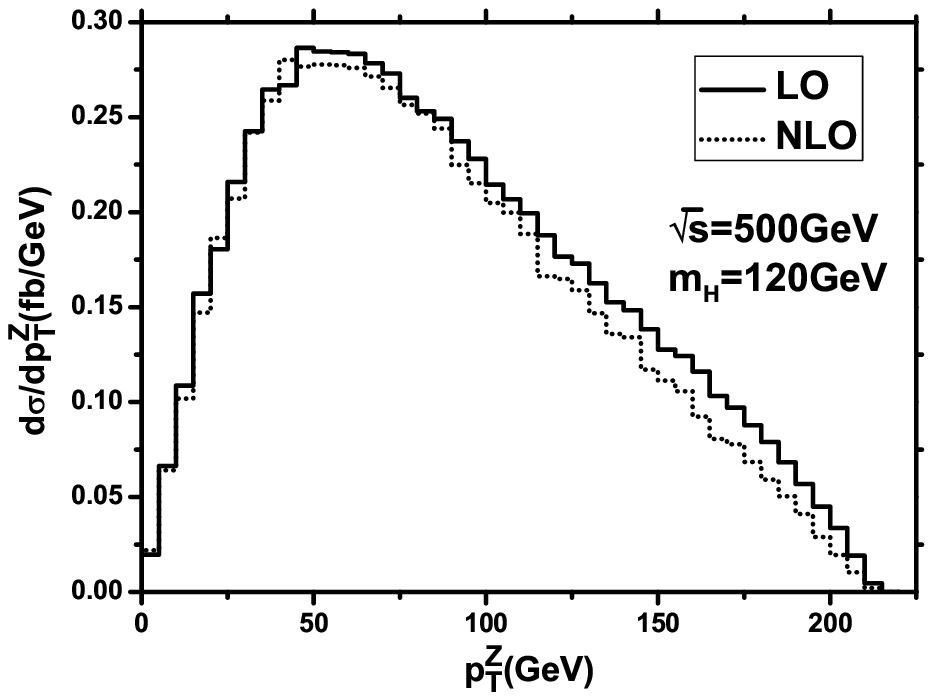}
\caption{\label{fig5} Distributions of the transverse momenta of
$W^-$- and $Z^0$-boson for the \eewwz process at the LO and EW NLO
with $\sqrt{s}=500~GeV$ and $m_H=120~GeV$. (a) for $W^-$-boson,
(b) for $Z^0$-boson. }
\end{figure}

Because the distribution of transverse momenta of $p_T^{W^-}$ is
the same as that of $p_T^{W^+}$ in the CP-conserving SM, we
provide only the distributions of $p_T^{W^-}$ at the LO and EW NLO
in Fig.\ref{fig5}(a). The differential cross sections of
transverse momentum of $Z^0$-boson at the LO and up to NLO
($\frac{d\sigma_{LO}}{dp_T^{Z}}$ and
$\frac{d\sigma_{NLO}}{dp_T^{Z}}$) are drawn in Fig.\ref{fig5}(b).
In these two figures we take $m_H=120~GeV$ and $\sqrt{s}=500~GeV$.
We can see from Figs.\ref{fig5}(a,b) that the EW NLO corrections
generally suppress the LO differential cross sections especially
when $p_T^{W^-}(p_T^{Z})>100~GeV$.

\par
\begin{figure}
\centering
\includegraphics[scale=0.75]{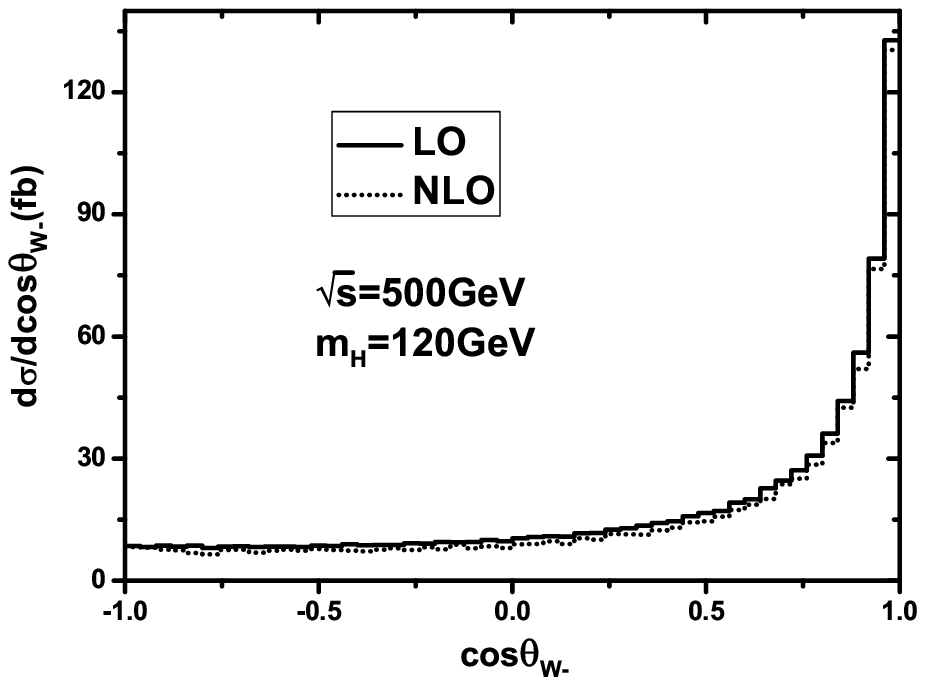}
\includegraphics[scale=0.75]{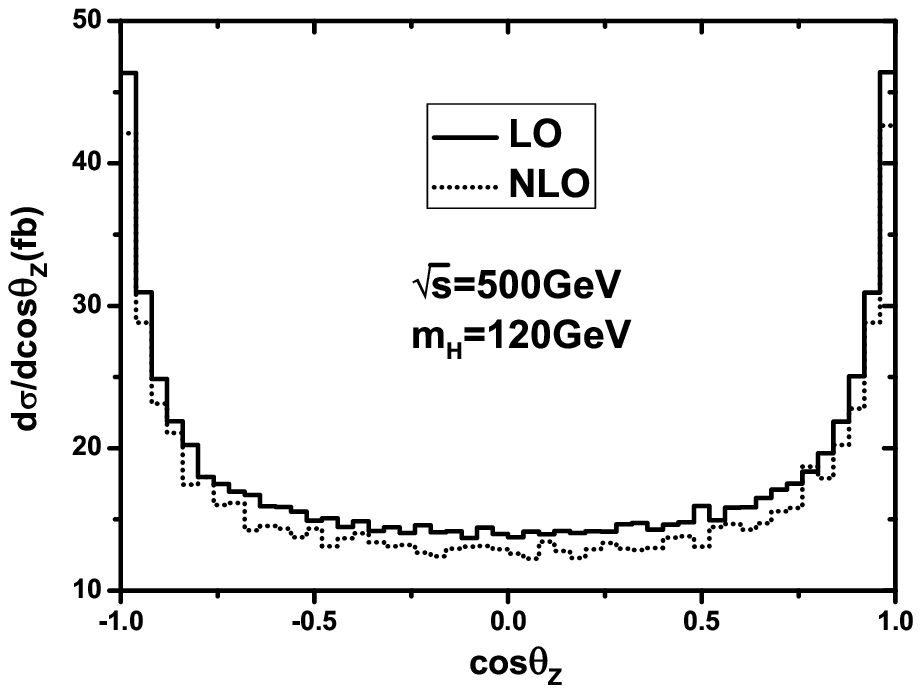}
\caption{\label{fig6} Distributions of the cosine of the
$W^-$-boson ($Z^0$-boson) production angle with respect to z-axis
for the \eewwz process at the LO and EW NLO with
$\sqrt{s}=500~GeV$ and $m_H=120~GeV$. (a) for
$\frac{d\sigma_{LO,NLO}}{d \cos\theta_{W^-}}$, (b) for
$\frac{d\sigma_{LO,NLO}}{d \cos\theta_{Z}}$. }
\end{figure}

We take the orientation of incoming electron as the z-axis. The
$\theta_{W^-}$ (or $\theta_{Z}$) is defined as the $W^-$-boson (or
$Z^0$-boson) production angle with respect to the z-axis. In
Figs.\ref{fig6}(a,b) we present the LO and EW NLO distributions of
cosines of the pole angles of $W^-$- and
$Z^0$-boson($\cos\theta_{W^-}$ and $\cos\theta_{Z}$) respectively,
in conditions of $\sqrt{s}=500~GeV$ and $m_H=120~GeV$. Both LO and
NLO curves in Fig.\ref{fig6}(a) show that the produced $W^-$-boson
declines to go out in the forward hemisphere, while
Fig.\ref{fig6}(b) demonstrates that the LO and NLO distributions
of the outgoing $Z^0$-boson are symmetry in the forward and
backward hemisphere regions.

\par
\begin{figure}
\centering
\includegraphics[scale=0.74]{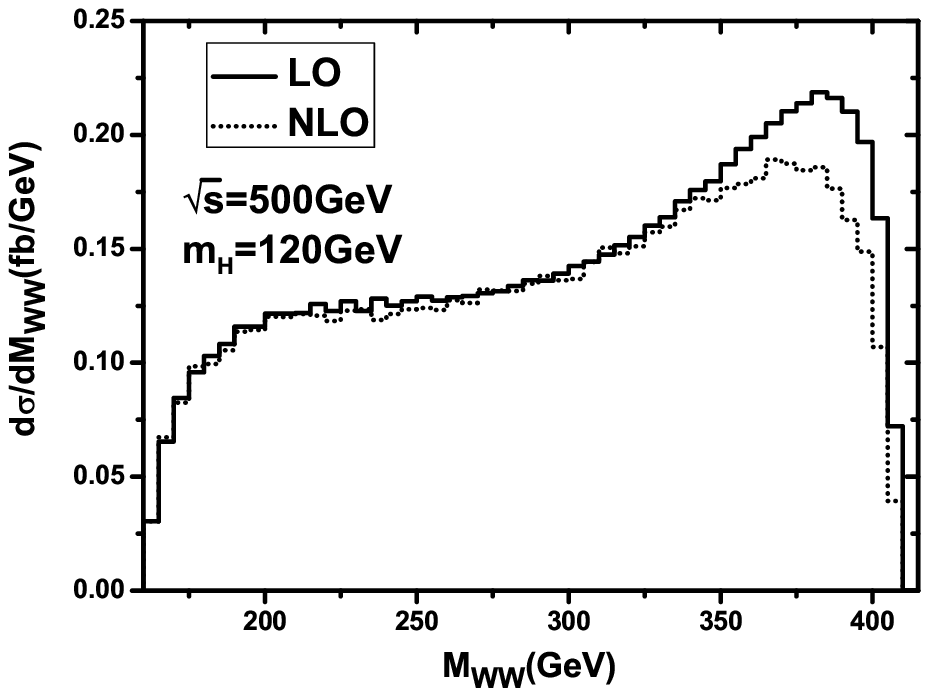}
\includegraphics[scale=0.74]{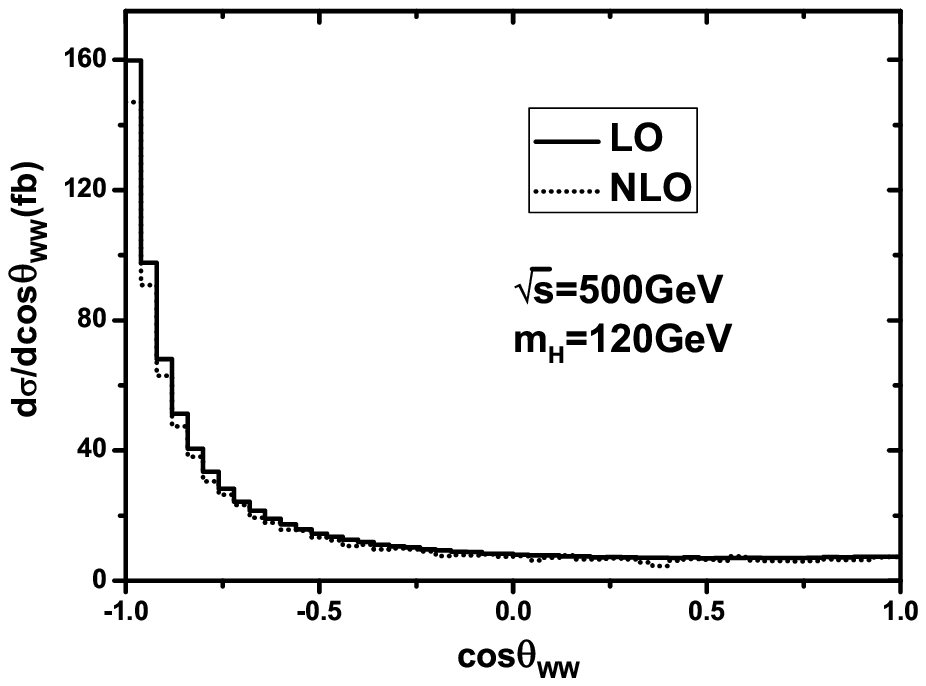}
\caption{\label{fig7} (a) Distributions of the invariant mass of
$W$-pair at the LO and EW NLO when $m_H=120~GeV$ and
$\sqrt{s}=500~GeV$. (b) Differential cross sections of the cosine
of the angle between the produced $W$-pair at the LO and EW NLO
with $m_H=120~GeV$ and $\sqrt{s}=500~GeV$.}
\end{figure}

The distributions of the $W$-pair invariant mass $M_{WW}$ at the LO
and EW NLO are shown in Fig.\ref{fig7}(a), and the differential
cross sections of the cosine of the angle between the produced
$W$-pair at the LO and EW NLO are presented in Fig.\ref{fig7}(b)
where we take $m_H=120~GeV$ and $\sqrt{s}=500~GeV$. We can see from
the Fig.\ref{fig7}(a) that there is an enhancement in the relatively
large $M_{WW}$ region(from $350~GeV$ to $400~GeV$) for each of the
LO and NLO distributions, and the EW NLO correction suppresses
significantly the LO differential cross section
$d\sigma_{LO}/dM_{WW}$ in this region. Fig.\ref{fig7}(b) shows the
LO and EW NLO distributions of cosine of the angle between the
produced $W$-pair. And we can see from the figure that the produced
$W$-pair prefer to go out almost back to back, that leads to the
$M_{WW}$ having the tendency to distribute in large value region.
That is why we see an enhancement in the large $M_{WW}$ region for
each of the LO and NLO distributions as shown in Fig.\ref{fig7}(a).

\par
On the analogy of the definitions of the LO and NLO
forward-backward charge asymmetry for top-quark in
Ref.\cite{Wein}, we define the LO and NLO forward-backward charge
asymmetries of $W^-$-boson as,
\begin{eqnarray}
\label{chargeAS}
A_{FB,LO}^{W^-}=\frac{\sigma_{tree}^-}{\sigma_{tree}^+},~~~~~
A_{FB,NLO}^{W^-}=\frac{\sigma_{tree}^-}{\sigma_{tree}^+}\left(1+\frac{\Delta
\sigma_{tot}^-}{\sigma_{tree}^-}-\frac{\Delta
\sigma_{tot}^+}{\sigma_{tree}^+}\right ).
\end{eqnarray}
The explicit expressions for $\sigma_{tree}^{\pm}$ are defined as
\begin{eqnarray}
\sigma_{tree}^{\pm}=\sigma_{tree}(y_{W^-}>0)\pm\sigma_{tree}(y_{W^-}<0),
\end{eqnarray}
where $y_{W^-}$ is the rapidity of $W^-$-boson, the notations
$\sigma_{tree}(y_{W^-}>0)$ and $\sigma_{tree}(y_{W^-}<0)$
represent the cross sections for the produced $W^-$-bosons in the
forward and backward hemispheres at the LO respectively. The
forward direction is along the orientation of z-axis.
$\Delta\sigma_{tot}^{\pm}$ denote the EW NLO corrections to the
cross sections $\sigma^{\pm}_{tree}$. In the conditions of
$\sqrt{s}=500~GeV$ and $m_H=120~GeV$, we get
$A_{FB,LO}^{W^-}=50.98(2)\%$ and $A_{FB,NLO}^{W^-}=53.28(3)\%$.
Both the LO and NLO results show that most of the $W^-$-bosons are
produced in the forward hemisphere, that feature has been already
demonstrated in Fig.\ref{fig6}(a).

\vskip 8mm
\section{Summary}
The $W^+W^-Z^0$ production via electron-positron collision at the
ILC is an important process not only in probing the non-Abelian
structures of the SM, but also in finding new physics. In this
report we have shown that the phenomenological effects due to the
one-loop EW radiative corrections, can be demonstrated in the
\eewwz process for all colliding energies ranging from $300~GeV$
to $1~TeV$ at the ILC. Our results show the EW one-loop radiative
corrections significantly suppress the LO cross sections, and the
relative correction to the cross section varies from $-17.6\%$ to
$-5.3\%$ when $m_H = 120~GeV$ and $\sqrt{s}$ goes up from
$300~GeV$ to $1~TeV$. We can see the obvious effects of the EW NLO
correction on the physical observables, such as, the distributions
of the transverse momenta of final $Z^0$- and $W$-bosons, the
differential cross section of the invariant mass of $W$-pair, the
distribution of the angle between $W$-pair, the production pole
angle distributions of $W^-$- and $Z^0$-boson, and the
forward-backward charge asymmetry of $W^-$-boson.

\vskip 8mm
\par
\noindent{\large\bf Acknowledgments:} This work was supported in
part by the National Natural Science Foundation of
China(No.10875112, No.10675110), the National Science Fund for
Fostering Talents in Basic Science(No.J0630319).


\vskip 8mm
\begin{thebibliography}{99}
\bibitem{s1}
S. L. Glashow, Nucl. Phys. {\bf 22} (1961) 579; S. Weinberg, Phys.
Rev. Lett. {\bf 1} (1967) 1264; A. Salam, Proc. 8th Nobel
Symposium Stockholm 1968,ed. N. Svartholm (Almquist and Wiksells,
Stockholm 1968) p.367; H. D. Politzer, Phys. Rep. {\bf 14} (1974)
129.

\bibitem{s2}
P. W. Higgs, Phys. Lett {\bf 12} (1964) 132, Phys. Rev. Lett. {\bf
13} (1964) 508; Phys. Rev. {\bf 145} (1966) 1156; F. Englert and
R.Brout, Phys. Rev. Lett. {\bf 13} (1964) 321; G. S. Guralnik, C.
R. Hagen and T. W. B. Kibble, Phys. Rev. Lett. {\bf 13} (1964)
585; T. W. B. Kibble, Phys. Rev. {\bf 155} (1967) 1554.

\bibitem{measurement of charged TGC at LEP2}
G. Gounaris {\it et al.}, in {\it Physics at LEP 2}, Report CERN
96-01 (1996), eds G. Altarelli, T. Sj\"{o}strand, F. Zwirner, Vol.
1, 525.

\bibitem{hep-ex/0412015}
The LEP Collaborations, {\it A combination of preliminary
electroweak measurements and constraints on the standard model},
CERN-PH-EP/2004-069, arXiv: hep-ex/0412015v2.

\bibitem{part 1}
M. Lemoine and M. J. G. Veltman, Nucl. Phys. {\bf B164} (1980)
445; M. Bohm, A. Denner, T. Sack, W. Beenakker, F. A. Berends and
H. Kuijf, Nucl. Phys. {\bf B304} (1988) 463; J. Fleischer, F.
Jegerlehner and M. Zralek, Z. Phys. {\bf C42} (1989) 409; W.
Beenakker, A. Denner, S. Dittmaier, R. Mertig and T. Sack, Nucl.
Phys. {\bf B410} (1993) 245; A. Denner, S. Dittmaier, M. Roth and
L. H. Wieders, Phys. Lett. {\bf B612} (2005) 223; Nucl. Phys. {\bf
B724} (2005) 247.

\bibitem{part 2}
W. Beenakker, F. A. Berends and A. P. Chapovsky, Nucl. Phys. {\bf
B548} (1999) 3.

\bibitem{2-loop corrections to W-pair production}
J. H. K\"{u}hn, F. Metzler and A. A. Penin, Nucl. Phys. {\bf B795}
(2008) 277.

\bibitem{diboson production at the Tevatron}
Mark S. Neubauer, FERMILAB-CONF-06-115-E, arXiv: hep-ex/0605066v2;
Junjie Zhu, arXiv: 0907.3239v1; The D0 Collaboration, V. Abazov,
{\it et al.}, {\it Combined measurements of anomalous charged
trilinear gauge-boson couplings from diboson production in
$p\bar{p}$ collisions at $\sqrt{s}=1.96$ TeV},
FERMILAB-PUB-09-380-E, arXiv: 0907.4952v1.

\bibitem{VVV production at hadron colliders}
A. Lazopoulos, K. Melnikov and F. Petriello, Phys. Rev. {\bf D76}
(2007) 014001; V. Hankele and D. Zeppenfeld, Phys. Lett. {\bf
B661} (2008) 103; T. Binoth, G. Ossola, C. G. Papadopoulos and R.
Pittau, JHEP {\bf 0806} (2008) 082.

\bibitem{ILC}
Parameters for Linear Collider,
http://www.fnal.gov/directorate/icfa/LC\_parameters.pdf

\bibitem{sujijuan}
Su Ji-Juan, Ma Wen-Gan, Zhang Ren-You, Wang Shao-Ming and Guo Lei,
Phys. Rev. {\bf D78} (2008) 016007.

\bibitem{tree-level results for WWZ production}
V. Barger, T. Han and R. J. N. Phillips, Phys. Rev. {\bf D39}
(1989) 146; C. Grosse-Knetter and D. Schildknecht, Phys. Lett.
{\bf B302} (1993) 309; S. Dawson, A. Likhoded, G. Valencia and O.
Yushchenko, arXiv: hep-ph/9610299; T. Han, H. -J. He and C. -P.
Yuan, Phys. Lett. {\bf B422} (1998) 294; M. Beyer, S. Christ, E.
Schmidt and H. Schroeder, arXiv: hep-ph/0409305.

\bibitem{fey}
T. Hahn,  Comput. Phys. Commun. {\bf 140} (2001) 418.

\bibitem{formloop}
T. Hahn, M. Perez-Victoria, Comput. Phys. Commun. {\bf 118} (1999)
153.

\bibitem{Denner}
A. Denner, Fortschr. Phys. {\bf 41} (1993) 307.

\bibitem{DR}
G. 't Hooft and M. Veltman, Nucl. Phys. {\bf B44} (1972) 189.

\bibitem{COMS scheme}
D. A. Ross and J. C. Taylor, Nucl. Phys. {\bf B51} (1979) 25.

\bibitem{Passarino}
G. Passarino and M. Veltman, Nucl. Phys. {\bf B160} 151 (1979).

\bibitem{Five}
A. Denner and S. Dittmaier, Nucl. Phys. {\bf B658} (2003) 175.

\bibitem{OneTwoThree}
G.'t Hooft and M. Veltman, Nucl. Phys. {\bf B153} (1979) 365.

\bibitem{Four}
A. Denner, U Nierste and R Scharf, Nucl. Phys. {\bf B367} (1991)
637.

\bibitem{van}
G. J. van Oldenborgh, Comput. Phys. Commun {\bf 58} (1991) 1.

\bibitem{hepdata}
C. Amsler, et al., Phys. Lett. {\bf B667} (2008) 1.

\bibitem{KLN}
T. Kinoshita, J. Math. Phys. {\bf 3} (1962) 650; T. D. Lee and M.
Nauenberg, Phys. Rev. {\bf 133} (1964) 1549.

\bibitem{PSS}
B. W. Harris and J. F. Owens, Phys. Rev. {\bf D65} (2002) 094032.

\bibitem{leger}
F. Jegerlehner, Report No. DESY 01-029, arXiv: hep-ph/0105283.

\bibitem{Wein}
S. Dittmaier, P. Uwer and S. Weinzierl, Eur. Phys. J. {\bf C59}
(2009) 625.

\end{thebibliography}
\end{document}